\documentclass[conference]{IEEEtran}
\IEEEoverridecommandlockouts
\usepackage{cite}
\usepackage{amsmath,amssymb,amsfonts}
\usepackage{algorithmic}
\usepackage{graphicx}
\usepackage{textcomp}
\usepackage{xcolor}
\def\BibTeX{{\rm B\kern-.05em{\sc i\kern-.025em b}\kern-.08em
    T\kern-.1667em\lower.7ex\hbox{E}\kern-.125emX}}
\begin{document}

\title{Topological Estimation of Number of Sources \\ in Linear Monocomponent Mixtures}

\author{\IEEEauthorblockN{Sean Kennedy}
\IEEEauthorblockA{\textit{Electrical and Computer Engineering} \\
\textit{Naval Postgraduate School}\\
Monterey, CA, USA \\
sean.kennedy@nps.edu}
\and
\IEEEauthorblockN{Murali Tummala}
\IEEEauthorblockA{\textit{Electrical and Computer Engineering} \\
\textit{Naval Postgraduate School}\\
Monterey, CA, USA \\
mtummala@nps.edu}
\and
\IEEEauthorblockN{John McEachen}
\IEEEauthorblockA{\textit{Electrical and Computer Engineering} \\
\textit{Naval Postgraduate School}\\
Monterey, CA, USA \\
mceachen@nps.edu}
}

\maketitle

\begin{abstract}
Estimation of the number of sources in a linear mixture is a critical preprocessing step in the separation and analysis of the sources for many applications.  Historically, statistical methods, such as the minimum description length and Akaike information criterion, have been used to estimate the number of sources based on the autocorrelation matrix of the received mixture.  In this paper, we introduce an alternative, topology-based method to compute the number of source signals present in a linear mixture for the class of constant-amplitude, monocomponent source signals. As a proof-of-concept, we include an example of three such source signals that overlap at multiple points in time and frequency, which the method correctly identifies from a set of eight redundant measurements.  These preliminary results are promising and encourage further investigation into applications of topological data analysis to signal processing problems. 
\end{abstract}


\section{Introduction} \label{sec:Introduction}
The objective of Blind Source Separation (BSS), also called blind signal separation or source separation, is to separate a group of source signals from a mixture without detailed knowledge of the sources or mixing process itself. BSS has applicability to many radar, communication, and imaging scenarios. Numerous techniques have been developed for BSS of mixed signals~\cite{BSSreview}. However, in most cases, the separation techniques assume that the number of sources equals the number of observations or that the number of sources is known in advance~\cite{BSSreview,MUSIC,ESPRIT}.  Thus, the ability to estimate or determine the number of sources to be unmixed is a critical preprocessing step in practical implementation.

The problem of estimating the number of sources in a linear mixture has been studied in the literature~\cite{ZhaoDetectNumSourcesInWhiteNoise,FishlerNumSources,salmanEstimatingNumberSources2015,WeissBlindNumSources}.  The two most popular methods used for this estimation are information-theoric measures based on the statistics of the mixture's autocorrelation matrix: the Minimum Description Length (MDL) and Akaike Information Criterion (AIC) estimators~\cite{FishlerNumSources,salmanEstimatingNumberSources2015}. Both have been shown to work well under the assumption of temporally and spatially white noise and Gaussian-distributed random sources. However, they have also been shown to be non-robust when real-world data deviates from these source and noise models~\cite{salmanEstimatingNumberSources2015}.  While numerous modifications and enhancements to these methods have been proposed, e.g.,~\cite{FishlerNumSources,salmanEstimatingNumberSources2015,WeissBlindNumSources}, the fundamental underpinning of these methods remains rooted in the analysis of eigenvalues and eigenvectors of the sample autocorrelation matrix with the assumption that there exist fewer sources than mixed measurements.  

In this paper, we introduce an approach that departs from the statistical inference model for estimating the number of sources.  Instead, we frame the problem in terms of topology and show that existing Topological Data Analysis (TDA) tools can be used to estimate the number of independent sources under certain scenarios.  The method is mathematically motivated and developed in Section~\ref{sec:method}. In Section~\ref{sec:Example}, we provide a simple validation of the method for three nonstationary sources. Conclusions and a discussion of future research goals are provided in Section~\ref{sec:conclusions}.

\section{Method Development and Discussion} \label{sec:method}

In practical terms, the method consists of only three steps: (1) embed an observed signal as a manifold in higher dimensional space, (2) compute the Betti number sequence~\cite{Betti} of the manifold using TDA, and (3) match the Betti number sequence to a known reference sequence. As shown in Section~\ref{sec:Example}, implementation of these steps is straightforward with appropriate software.  Thus, the primary contribution of this paper is the analysis of the mathematical mechanisms by which the method achieves valid results.  In Subsection~\ref{sec:TDA_mono} we provide the basic topological theory behind the method.  In Subsection~\ref{sec:embedding} we apply this theory to the case of a linear mixture as might be encountered in a radar, sonar, or multi-receiver communication array.  In Subsection~\ref{sec:independence}, we discuss the primary constraint the measured data must meet in order for the method to work, and provide a potential avenue to circumvent the constraint in practice.  In Subsection~\ref{sec:topology_comp}, we briefly describe the tool used to compute the Betti number sequence of a given data set, and how to use this sequence to estimate the number of sources in the mixture.

\subsection{Topological Analysis of Monocomponent Mixtures} \label{sec:TDA_mono}

We begin by analyzing the mixing problem through the lens of topology.  First, let $x(t) = [x_1(t), x_2(t),...,x_n(t)]^{\top}$ be the vector of $n$ independent sources. If we consider each $x_i(t)$ as the motion of a point along an orthogonal basis vector of $\mathbb{R}^n$, then $x(t)$ can be interpreted as a parametric path through $\mathbb{R}^n$. We restrict each $x_i(t)$ to be a continuous signal of the form $x_i(t) = A_i \textrm{cos}(\alpha_i(t))$, where $A_i$ is the constant amplitude of $x_i(t)$, and $\alpha_i(t)$ is a continuous function of time encoding the instantaneous frequency and phase of $x_i(t)$. Since each $\alpha_i$ could include a constant phase term in the interval $[-\pi,\pi]$, we can say that these sinusoidal sources are ``cosines" without any loss of generality. Sources of this type are often referred to as constant-amplitude ``monocomponent signals'' in the literature and are frequently encountered in radar (e.g., chirps) and telecommunication (e.g., continuous phase frequency modulated signals) applications.

Constant-amplitude monocomponent signals are of special interest since they can be embedded as topological circles in $\mathbb{R}^2$~\cite{Kennedy}.  To see this, consider that the Hilbert transform of $x_i(t)$ is immediately obtained as $\widetilde{x}_i(t) = A_i \textrm{sin}(\alpha_i(t))$, where $A_i$ and $\alpha_i$ are unchanged from their $x_i(t)$ counterparts~\cite{HilbertTransform} . Then, by simple trigonometric identity, the expression $x_i^2(t)+\widetilde{x}_i^2(t)$ equals the constant $A_i^2$ for all times $t$, which is the definition of a circle in the plane~\cite{Kennedy}. Considering now each component $x_i(t)$ and its Hilbert transform $\widetilde{x}_i(t)$ as the motion of a point along mutually orthogonal axes in $\mathbb{R}^{2n}$, we find that the trajectory of this point (i.e., the \textit{phase portrait} of $x(t)$) forms a path on an $n$-torus since $(\mathbb{S}^1)^n=\mathbb{T}^n$~\cite{Recurrent}.  As discussed in~\cite{Recurrent}, the path itself may not actually become dense on the torus but form a torus knot instead when the ratio of instantaneous frequencies of individual components of $x(t)$ are rational throughout the observation window. This condition is unlikely to occur for incoherent signals with nonstationary  frequencies, so we set aside this outlier case and assume that the path becomes dense on the $\mathbb{T}^n$ manifold. 

By the K\"{u}nneth formula~\cite{Kunneth}, the Betti number sequence for an $n$-torus is given by the coefficients of the Poincar\'e polynomial $(1+q)^n$.  This sequence is a topological invariant of the $n$-torus, i.e., it is invariant under homeomorphisms and embeddings~\cite{Betti}. The existence of this topological invariant allows us to compute the number of sources in a mixture of monocomponent signals as follows. Let $y(t) \in \mathbb{R}^{m}$ be the vector composed of $m$ independent observations of mixtures of $x(t)$, and let $z(t) \in \mathbb{R}^{2n}$ be defined as $[x_1(t), \widetilde{x}_1(t),x_2(t), \widetilde{x}_2(t),...,x_n(t), \widetilde{x}_n(t)]^{\top}$. Let $w(t) \in \mathbb{R}^{2m}$ be a vector defined by $w(t) = f( z(t))$, where $f$ is an arbitrary function mapping $\mathbb{R}^{2n} \to \mathbb{R}^{2m}$.  When $f$ is a homeomorphism (when $m=n$) or an embedding (when $m>n$), the Betti number sequences of the phase portraits of $w(t)$ and $z(t)$ will be equal.  So, if we can find a suitable method to embed the observed vector $y(t)$ into $\mathbb{R}^{2m}$ as $w(t)$ such that $f$ exists and is a homeomorphism or embedding, then we can simply compute the Betti number sequence of the phase portrait of $w(t)$ to recover the Betti number sequence of the phase portrait of $z(t)$.  Subsection~\ref{sec:embedding} provides such a method for the scenario of a generic receiver array.  Once computed, if we find that the Betti number sequence of $z(t)$ matches the the coefficients of the polynomial $(1+q)^n$, we know that there exist $n$ monocomponent sources in the mixture.  In summary, our estimation strategy is: (1) embed an $\mathbb{R}^{m}$ signal mixture into $\mathbb{R}^{2m}$, (2) compute the Betti number sequence, and (3) compare the sequence to the coefficients of $(1+q)^n$.

\subsection{Embedding Observations into $\mathbb{R}^{2m}$} \label{sec:embedding}

In order to identify a suitable embedding of $y(t)$ into $\mathbb{R}^{2m}$ as $w(t)$, we impose a constraint on the mixing process itself: that each observation vector $y_i(t)$ is given by:
\begin{equation} \label{eq:yi_orig}
    y_i(t) = \sum_{j=1}^{n} B_{i,j} \, \textrm{cos}(\alpha_j(t)+\phi_{i,j})
\end{equation}
for some values $B_{i,j} > 0$ and $\phi_{i,j} \in [-\pi,\pi]$.  In other words, each observation vector $y_i(t)$ is a sum of each of the sources of $x(t)$, modified by a relative magnitude and phase at each measurement.  This model was chosen as a simplification of the case of a receiver array with multiple, directional elements receiving mixtures of the source signals, as might be encounted in radar, sonar, or communications applications.  The directionality of the receiver elements along with the physical spacing of the source signals induces differences in magnitude among the components of each mixture.  Likewise, the physical spacing of the receiver's antenna induces a small time delay between the reception of different sources, which, if small, can be approximated by a phase-shift~\cite{ArrayProcessingText}.

An effective way to ensure that that the map $f: \mathbb{R}^{2n} \to \mathbb{R}^{2m}$ ($m \ge n$) exists and is a homeomorphism (or embedding) is to let $f$ be defined by a matrix $T$ such that $w(t) = Tz(t)$.  Then, $T$ is a linear map, which is guaranteed to be a homeomorphism (or embedding) if $T$ has full column rank of $2n$.  Letting $B_{i,j} = R_{i,j} \cdot A_i$ for some $R_{i,j}$, and using the trigonometric identity for angle sums, we rewrite $y_i(t)$ as:
\begin{equation} \label{eq:yi}
    y_i(t) = \sum_{j=1}^{n} R_{i,j} \, \textrm{cos}(\phi_{i,j})\,x_j(t) - R_{i,j} \,\textrm{sin}(\phi_{i,j})\, \widetilde{x}_j(t)
\end{equation}
where $\widetilde{x}_j(t)$ is the Hilbert transform of $x_j(t)$.  Based on the form of Eq.~\ref{eq:yi}, we can represent each component of $y(t)$ as a linear combination of the components of $z(t)$: the coefficient terms $R_{i,j}\,\textrm{cos}(\phi_{i,j})$ and $R_{i,j}\,\textrm{sin}(\phi_{i,j})$ form the entries of $T$.  By taking the Hilbert transform of each $y_i(t)$, we can immediately obtain a second set of $m$ observations given by:
\begin{equation}  \label{eq:ybi}
    \widetilde{y}_i(t) = \sum_{j=1}^{n} R_{i,j} \, \textrm{sin}(\phi_{i,j})\,x_j(t) + R_{i,j} \,\textrm{cos}(\phi_{i,j})\, \widetilde{x}_j(t)
\end{equation}
where each $R_{i,j}$ and $\phi_{i,j}$ are unchanged between $y_i(t)$ and its Hilbert transform $\widetilde{y}_i(t)$.  This second set of $m$ observations is likewise a linear combination of the components of $z(t)$, providing the additional $m$ rows of $T$.  Therefore, $w(t)$ is defined as $[y_i(t),\widetilde{y}_i(t)]^{\top}$, and $w(t) = Tz(t)$.  We note that in practice the embedding observation vector $\widetilde{y}_i(t)$ can be obtained either through direct measurement (e.g., analog phase shifter circuitry), or through digital analysis via the discrete Hilbert transform.

A useful result of requiring that the map $f: \mathbb{R}^{2n} \to \mathbb{R}^{2m}$ $(m\ge n)$ be a linear transformation $T$ is that if $T$ exists and is full rank, then there also must exist a pseudo-inverse of $T$ that recovers the source vectors from the observation vectors.  Therefore when the method is successful, in addition to the mere assertion that there are $n$ sources in the mixture, we can further assert that the $n$ sources are \textit{able to be unmixed} by finding the appropriate pseudoinverse, e.g., via a least squares fit of transformed data to the $n$-dimensional square torus.  This is a powerful insight that will be explored in future work.

Another potential benefit of this method is that the only constraint on $T$ is that $T$ be full rank, and so exact knowledge of the entries of $T$ (i.e., each $B_{i,j}$ and $\phi_{i,j}$) is unnecessary.  In practical terms, this means that prior knowledge of the array geometry, incidence angle, propagation velocity, or wave frequencies are generally not required for the method to work. This could allow the method to be used on mobile, distributed, and/or dynamic arrays of receiver elements.  

\subsection{Defining Independence in the Observations} \label{sec:independence}

As explained in Section~\ref{sec:embedding}, $T$ must have full column rank under this ``Hilbert embedding'' for our topological feature recovery strategy to work, so we now discuss the conditions under which this is true.  Label each row of $T$ corresponding to the equation $y_i(t)$ as $r_i(t)$ and each row of $T$ corresponding to the equation $\widetilde{y}_i(t)$ as $\widetilde{r}_i(t)$.  Then, each pair of rows $[r_i(t),\widetilde{r}_i(t)]^\top$ forms $n$ blocks of size $2 \times 2$ of the form:
\begin{equation}  \label{twoByTwo}
    \begin{bmatrix} R_{i,j}\cos(\phi_{i,j}) & -R_{i,j}\sin(\phi_{i,j}) \\
                    R_{i,j}\sin(\phi_{i,j}) &  R_{i,j}\cos(\phi_{i,j})
    \end{bmatrix}
\end{equation}
where the terms have the same meanings as in Eqs.~\ref{eq:yi} and~\ref{eq:ybi}. This is the familiar form of a matrix in $\mathbb{C}^{m \times n}$ as implemented in $\mathbb{R}^{2m \times 2n}$.  As such, we construct a dual of the matrix $T$, called $U$, where each complex term of $U$ replaces each $2 \times 2$ block from Eq.~\ref{twoByTwo} with $U_{i,j} = R_{i,j} \angle{\phi_{i,j}}$ (in phasor notation).  Under this construction, we can say that $T$ has full column rank of $2n$ if and only if $U$ has full column rank $n$. Labeling each row of $U$ as $u_i$ for $i \in [1,m]$, we note that each $u_i$ corresponds to the coefficients of the complex observation vector $v_i(t) = y_i(t) + \sqrt{-1} \, \widetilde{y}_i(t)$.  Then, for $U$ to possess rank $n$, there must exist a subset of $n$ rows chosen from $u_i$ that are linearly independent. Since the effect of multiplying the complex vector $v_k(t)$ by a complex constant $c_k = A_k \angle{\theta_k}$ ($A_k \in \mathbb{R}$, $\theta_k \in [-\pi,\pi]$) is to multiply the magnitudes of the associated $y_i(t)$ and $\widetilde{y}_i(t)$ vectors by $A_k$ and add $\theta_k$ to the phases of $y_i(t)$ and $\widetilde{y}_i(t)$, we can informally say that a set of observations $y(t)$ are ``independent'' if no element $y_i(t)$ is a linear combination (over $\mathbb{R}$) of any phase-shifted versions of the others.  Consequently, it is the relative ratios in the magnitudes and relative differences in the phases of each component among the $y_i(t)$ observations that determine the rank of $T$.

We suspect that the condition for a full rank $T$ will hold in most practical instances where $m \gg n$ (e.g., phased array with many elements) as only $n$ of the $m$ observation vectors must be ``independent'' in the meaning given above.  However, in cases where there are not expected to be $n$ independent observation vectors available (e.g., $m < n$), it may be possible to derive additional independent observation vectors through the use of a Time Delay Embedding (TDE), as discussed in~\cite{Recurrent}.  A full discussion of the theory, usage and drawbacks of TDEs as embedding functions is beyond the scope of this paper; the interested reader is referred to references~\cite{Recurrent,Takens,Kennedy,uncorrelated}.  With respect to monocomponent sources, TDEs have the potential to induce relative phase shifts among the source components in a given mixture~\cite{Kennedy}, and are therefore a potential source for additional independent measurement vectors when there are fewer array elements than sources present in the mixture.

\subsection{Topological Computation via Persistent Homology} \label{sec:topology_comp}

As discussed in Sections~\ref{sec:TDA_mono} and~\ref{sec:embedding}, once we have obtained a properly embedded analysis signal $w(t)$, our next step is to compute the Betti number sequence corresponding to the topology of the phase portrait of $w(t)$.  The TDA tool we choose to perform this computation with is Persistent Homology (PH)~\cite{Carlsson}, which has been used to study the topology of datasets with many real-world applications~\cite{periodEst,Wheeze,SW1PERS,PretermBirth}.  We direct the unfamiliar reader to reference~\cite{PHsurvey} for an introduction to the theory and computation of PH. Many PH computation packages have been developed in recent years~\cite{ComputingHomology}; in Section~\ref{sec:Example}, we use the JavaPlex software to compute PH due to its simple integration into MATLAB~\cite{Javaplex}.

Treating the PH computation as a ``black box'' processing step, we can estimate the Betti number sequence of the input manifold through simple analysis of the output ``barcode'' plot~\cite{Javaplex}.  For the purposes of this paper, we determine the Betti number sequence by simply counting the number of features in the barcode plot that persist for at least half of the PH computation interval in each dimension.  We use this sequence to compute the number of sources by finding an integer value $n$ for which our sequence matches the coefficients of the polynomial $(1+q)^n$, as discussed in Section~\ref{sec:TDA_mono}. This integer value, $n$, is the output of the method, providing the estimate of the number of sources in the mixture.

In practice, we need not actually search through all possible $n$ to find a match. As the coefficients of $(1+q)^n$ follow from the binomial theorem, and ${n \choose 1} = n$, we let our initial ``guess'' of $n$ equal the second Betti number (corresponding to homology group $H_1$) determined from the barcode plot. We then verify that the remainder of Betti numbers from the sequence match the remaining coefficients of $(1+q)^n$.  If they all match, we have confidence in our computation of the topology of $w(t)$ and thus $z(t)$, so we assert that there are $n$ sources to be unmixed.  If they do not match, then the Betti numbers do not conform to those of a torus $\mathbb{T}^n$, as there are no other possible values of $n$ which could produce the obtained Betti number sequence. We assume in the non-matching case that our computation has failed and we cannot accurately estimate the number of sources; this can occur, for example, when the mixing matrix $T$ is either singular or otherwise badly conditioned.

\section{Example Computation with Three Monocomponent Sources} \label{sec:Example}

\begin{figure*}[t!]
    \centering
    \includegraphics[width=\textwidth]{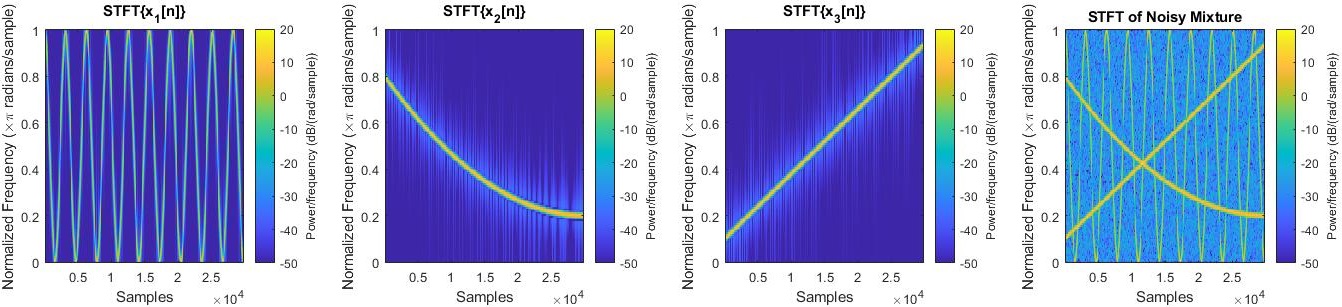}
    \caption{Short-time Fourier Transform (STFT) Spectrograms of clean source signals, along with one example of a mixture of all three sources with AWGN at 20 dB SNR.  As can be seen, all three signals are highly nonstationary causing overlap in time and frequency at various points.}
    \label{fig:sources}
\end{figure*}

We now present a proof-of-concept demonstration of the method using three synthetic digital signals in MATLAB.  While unknown to the receiver, the received signals are a mixture of three independent monocomponent sources with nonstationary frequency characteristics as could be encountered in a radar scenario. The first source is a ``barrage jammer'' waveform whose instantaneous frequency continuously sweeps the Nyquist range, while the other two are ``chirp'' signals that begin and end at disparate instantaneous frequencies.  Plots of the spectrograms (i.e., short-time Fourier transforms) of the three source signals, as well as an example of a noisy mixture of the three sources, are provided in Figure~\ref{fig:sources}.  These signals were chosen somewhat arbitrarily as an example of wideband, nonstationary signals that overlap at irregular points in the time and frequency domains and are therefore inseparable through classical linear filtering.  The specific parameters (e.g., frequency sweep ranges, initial phase, etc.) for these chosen signals are irrelevant; our topological method is general enough that virtually any form of monocomponent signals is viable under this method.  The authors have conducted many additional trials using alternate monocomponent signal types with results similar to those presented herein.

Assuming that the receiver array contains eight elements (eight being an arbitrary number substantially larger than the expected three sources), we generate eight random values of $R_{i,j} \in [0.75,1.25]$ and $\phi_{i,j} \in [-\pi,\pi]$ and create the eight observation signals, $y_i[n]$, according to a discretized version of Eq.~\ref{eq:yi_orig}.  Each observation vector contains 30000 samples, equivalent to a 1-MHz sampling frequency observed over 30 milliseconds. Additive white Gaussian noise (AWGN) is added separately to each observation signal with a signal-to-noise ratio (SNR) randomly selected from 15 to 25 dB.  A variable SNR on each measurement is used to provide some evidence of robustness in the presence of nonspatially white noise.  As discussed in Section~\ref{sec:embedding}, we compute the Hilbert transform of each of our observed $y_i[n]$ to obtain $\widetilde{y}_i[n]$. As a final preparatory step, we remove the first and last 10\% of samples from each $y_i[n]$ and $\widetilde{y}_i[n]$ to minimize the windowing effects of computing the Hilbert transform of a finite-duration signal.  These trimmed signals are combined into the 16-dimensional input signal $w[n]$ via concatenation.

\begin{figure}[tbh]
      \centering
      \includegraphics[width=\columnwidth]{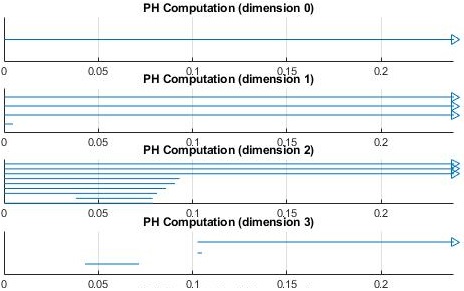}
      \caption{Barcodes corresponding to the PH computation in dimensions 0-3. The Betti numbers of $\{1,3,3,1\}$ are found by counting the number of lines in each dimension which persist over a majority of the filtration interval.}
      \label{fig:barcodes}
\end{figure}

We then provide the 16-dimensional signal $w[n]$ as the point-cloud input to JavaPlex.  We use the Witness Complex construction with 150 landmark points chosen using the sequential min-max procedure, up to a maximum filtration distance of 0.24~\cite{Javaplex}.  The results are provided in barcode format in Figure~\ref{fig:barcodes}, showing persistent Betti numbers of $\{1,3,3,1,0,0,0,0,0\}$.  As up to eight sources could be detected with this method, the PH was computed over dimensions 0-8; however, there were no features found in dimensions 4-8, so these empty results are omitted from Figure~\ref{fig:barcodes} to save space. It is trivial to identify that the computed Betti number sequence of $\{1,3,3,1\}$ matches the coefficients of the  polynomial $(1+q)^3 = 1 + 3q + 3q^2 + q^3$, thus the method computes that there are exactly 3 sources making up the mixed observation vectors, as was desired.  

For comparison purposes, we also computed the outputs of the MDL and AIC estimators as defined in~~\cite{salmanEstimatingNumberSources2015} on the same data set.  Surprisingly, these estimators improperly estimate that the mixtures consist of 7 independent sources.  Similar results are obtained when the parameters of the mixtures are changed and different noise levels are generated. As discussed in \cite{FishlerNumSources}, this error in the MDL and AIC estimators is likely due to variations in AWGN power among the observed signals.  While far from conclusive, this difference in outcomes hints that the topological method may provide additional robustness over the statistical methods in certain scenarios, and thus the method warrants additional research.

\section{Conclusions and Future Work} \label{sec:conclusions}

In this paper, we introduced a topology-based method to estimate the number of source signals present in a linear, monocomponent mixture. In particular, we showed how the analytic form of a mixture of $n$ monocomponent signals generates a phase space which is homeomorphic to the $n$-torus, so by using TDA tools to recover the topological features of the mixture, we can determine the number of signals making up the mixture.  While this method is only directly applicable to constant-amplitude monocomponent source signals in its current form, these signals are nonetheless commonplace in wireless communications and radar scenarios.  Therefore, the method could have substantial real-world applicability without significant modification.  

Our provided demonstration of the method is merely a proof-of-concept, as there are many aspects to the method that require further research in order to refine.  In particular, the TDA tool of PH is known to suffer from high computational complexity and sensitivity to noisy outliers in the data set~\cite{Javaplex}.  These issues are being addressed by other researchers in the literature~\cite{ComputingHomology}, but the nuances are complex.  Accordingly, we omitted such discussions in this paper as they distract from the presentation of the underlying theory of the technique.  Since we consider the PH computation as a ``black box'' for computing the Betti number sequence of the underlying mixture, any improvements in the robustness, accuracy, or efficiency of the PH computation can likewise be leveraged by our method.  We suspect additional preprocessing of the observation signals will also play a role in improving the method; these techniques will be explored in future work.

As briefly demonstrated in Section~\ref{sec:Example}, the method can potentially outperform the standard MDL and AIC methods in some specific scenarios, such as when the received noise is not spatially white.  While this outcome encourages further research into the method, a more robust investigation of this method is needed to adequately compare performance with the existing statistical methods.  Such a comparison requires additional control over the scenario model, any preprocessing and optimization steps, and variations in the numbers of sources, signal types, measurements, and noise types/levels.  Due to the large number of experimental parameters and alternate techniques to test, we have opted to save such an investigation for future publication, and present only the theory and proof-of-concept of our method here.

More generally, we believe the novelty of the approach is of sufficient interest: eschewing the usual tools of statistical analysis to instead analyze signals on the basis of shape.  As topological data analysis is a rapidly evolving field, we believe that additional practical algorithms and tools will emerge to tackle current problems in many fields such as signal processing, control theory, and communications.  As such, we intend to not only improve on this particular application, but also leverage the topological approach taken to search for novel solutions to other types of problems.  We strongly encourage other researchers to join us in this endeavour.

\vfill\pagebreak

\bibliographystyle{IEEE}
\bibliography{myRefs.bib}

\end{document}